\documentclass[aos,preprint]{imsart}
\RequirePackage[OT1]{fontenc}
\RequirePackage{amsthm,amsmath}
\RequirePackage[numbers]{natbib}
\RequirePackage[colorlinks,citecolor=blue,urlcolor=blue]{hyperref}
\usepackage{amssymb,cmmib57,graphics,color,mathrsfs,dsfont}
\usepackage{graphicx,epstopdf}
\usepackage{xcolor,soul}
\usepackage{pst-plot}
\usepackage{authblk}
\usepackage{algpseudocode}
\usepackage{multirow}
\usepackage{transparent}
\usepackage{todonotes}
\usepackage{amsmath}

\usepackage{algpseudocode} \usepackage{algorithm}

\usepackage{times}
\usepackage{bm}

 \usepackage{caption}
 
\usepackage{import}
\usepackage{xifthen}
\usepackage{pdfpages}
\usepackage{transparent}

 \usepackage{caption}
 
 \usepackage{pgfplots,tikz}
 \usetikzlibrary{shapes.geometric}

\usepackage{booktabs,amsfonts}
\usepackage{enumitem}

\def\p{\mathcal{P}}

\def\P{\text{pr}}

\def\P{\mathcal{P}}

\definecolor{darkgreen}{rgb}{0, .6, 0}

\arxiv{arXiv:0000.0000}

\startlocaldefs
\numberwithin{equation}{section}
\theoremstyle{plain}

\endlocaldefs

\begin{document}

{\bf\centering \large 
REAMP: A Stochastic Resonance Approach for Multi-Change Point Detection in High-Dimensional Data\\}

\noindent
{X. Shi\\}
{Department of Computer Science, Mathematics, Physics and Statistics,
University of British Columbia, Kelowna, BC V1V 1V7, Canada \texttt{xiaoping.shi@ubc.ca}\\}
{B. Jin\\}
{Department of Finance and Statistics, University of Science and Technology of China, Hefei, China \texttt{jbs@ustc.edu.cn}\\}
{X. Liu\\}
{School of Statistics, Jiangxi University of Finance and Economics, Nan Chang, China 330013 \texttt{liuxh109@126.com}\\}
{and Q. Li\\}
{Guangdong Provincial Key Laboratory of Interdisciplinary Research and Application; Data Science,  Beijing Normal-Hong Kong Baptist University, Zhuhai 519087, China \texttt{qiongli@bnbu.edu.cn}}

{\bf Abstract:} 
Detecting multiple structural breaks in high-dimensional data remains a challenge, particularly when changes occur in higher-order moments or within complex manifold structures. In this paper, we propose {REAMP} (Resonance-Enhanced Analysis of Multi-change Points), a novel framework that integrates optimal transport theory with the physical principles of stochastic resonance. By utilizing a two-stage dimension reduction via the Earth Mover's Distance (EMD) and Shortest Hamiltonian Paths (SHP), we map high-dimensional observations onto a graph-based count statistic. 
To overcome the locality constraints of traditional search algorithms, we implement a stochastic resonance system that utilizes randomized Beta-density priors to ``vibrate'' the objective function. This process allows multiple change points to resonate as global minima across iterative simulations, generating a candidate point cloud. A double-sharpening procedure is then applied to these candidates to pinpoint precise change point locations. We establish the asymptotic consistency of the resonance estimator and demonstrate through simulations that REAMP outperforms state-of-the-art methods, especially in scenarios involving simultaneous mean and variance shifts. The practical utility of the method is further validated through an application to time-lapse embryo monitoring, where REAMP provides both accurate detection and intuitive visualization of cell division stages.

\textbf{Keywords:} Stochastic Resonance; Earth Mover's Distance; Change Point Detection; High-Dimensional Data; Shortest Hamiltonian Path; Data Visualization.
 
\section{Introduction}
 
In the era of massive data availability, identifying structural changes in high-dimensional signals---such as medical image sequences, satellite imagery, or financial networks---remains a fundamental challenge in modern statistics and data science. This paper considers the canonical multiple change point model:
\begin{equation}\label{MCP}
F_i = \sum_{\ell=1}^{k^*+1} F_{G_\ell} I(\tau_{\ell-1} < i \leq \tau_\ell), \quad i = 1, \ldots, n,
\end{equation}
where $I(\cdot)$ is an indicator function, $F_i$ represents the distribution of a high-dimensional data matrix $(Y_i)_{s \times d}$ (e.g., pixel intensities in medical imaging), $\{F_{G_\ell}\}_{\ell=1}^{k^*+1}$ are $k^*+1$ distinguished distribution functions,  and $\tau_1, \ldots, \tau_{k^*}$ are the unknown change points to be estimated with $\tau_0=1$ and $\tau_{k^*+1}=n$. While model \eqref{MCP} appears simple, inferring structural breaks in high-dimensional settings is an arduous task, particularly when the dependencies within $Y_i$ require sophisticated time-series techniques or when the dimension $d$ grows faster than the sample size $n$.

Traditional high-dimensional change point detection (CPD) typically relies on cumulative sum (CUSUM) statistics, parametric modeling, or binary segmentation. For detecting mean-shifts, CUSUM-based methods for normal random vectors have been widely discussed \citep{Ji15, WS18}. However, without strong parametric assumptions, nonparametric methods like empirical divergence \citep{MJ14} or self-normalization \citep{WZVS22} are preferred. Recently, graph-based methods have attracted significant attention, as they allow for dimensionality reduction by projecting high-dimensional data into low-dimensional similarity structures while preserving the original manifold geometry \citep{CZ15, Sh24}.

Despite these advances, detecting \textit{multiple} change points remains difficult. Traditional CUSUM statistics often suffer from signal loss at localized peaks when multiple structural breaks are present \citep{FF22}. While methods like wild binary segmentation \citep{Fr14} or weighted CUSUM statistics with prior information \citep{SWR22} attempt to sharpen these searches, they often struggle with a ``centrality bias" or require heavy computational overhead. We propose a fundamentally different solution: \textbf{REAMP} (\underline{re}sonance-based \underline{amp}lification).

\subsection*{The REAMP Framework: Resonance and Amplification}
\begin{figure} [ht]
\centering
\makebox{\includegraphics[width=\textwidth]{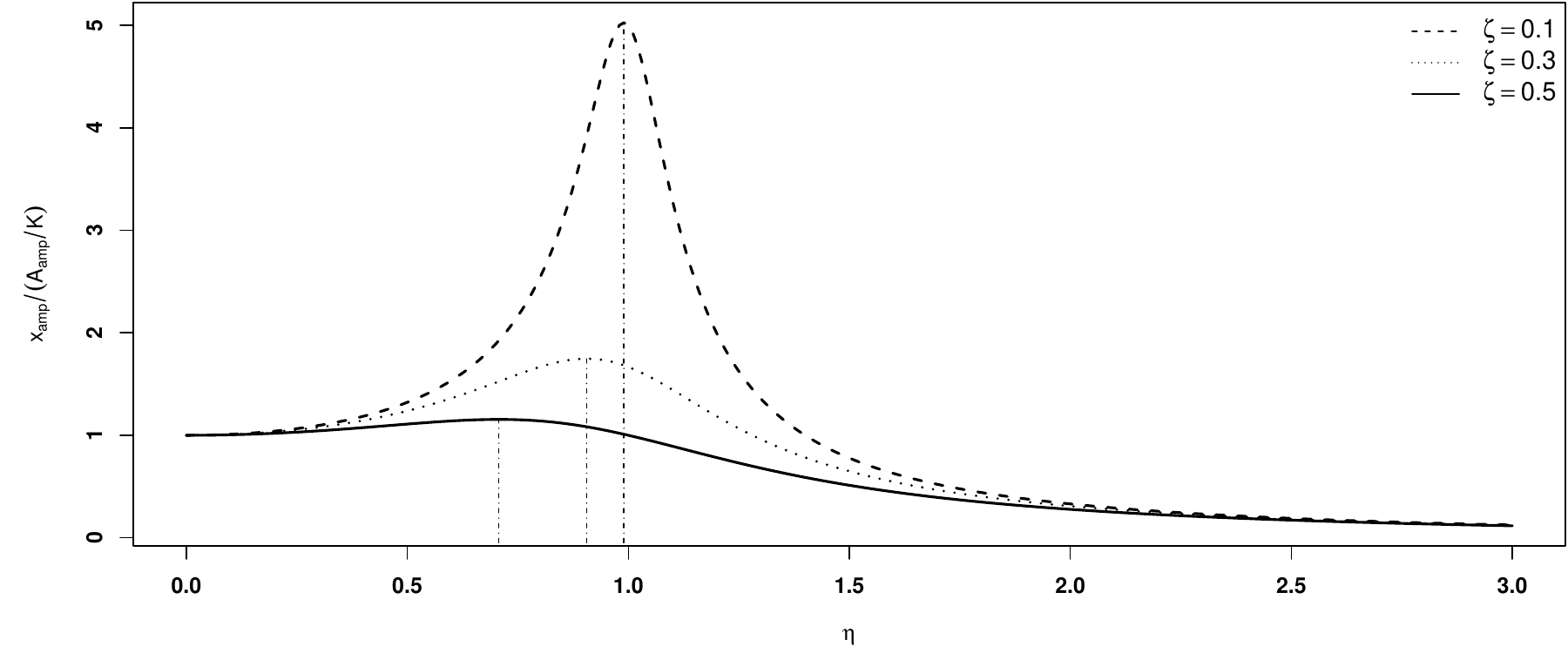}}
\caption{Plot of $x_{\text{amp}}/(A_{\text{amp}}/K)$ with maximum at $\eta=\sqrt{1-2\zeta^2}$.}
\label{fig-amp}
\end{figure}

Our technique draws inspiration from mechanical resonance, where an external source matches the vibration characteristics of a system to amplify a signal. In a mechanical system, the displacement amplitude is governed by the mass ($M$), damping ($C$), and stiffness ($K$):
\begin{equation}\label{amp}
x_{\text{amp}} = \frac{A_{\text{amp}}/K}{\sqrt{(1-\eta^2)^2 + (2\eta\zeta)^2}}
\end{equation}
where  $\eta=\omega/\sqrt{K/M}$ is the relative driving frequency with angular frequency $\omega$   and $\zeta=C/\sqrt{4KM}$ is the damping ratio   with viscous damping coefficient $C$". 
The resonant system above  {\it amplifies} the relative amplitude $A_{\text{amp}}/K$ by allowing $\eta$ to approach $\sqrt{1-2\zeta^2}$.   Fig \ref{fig-amp} plots
$x_{\text{amp}}/(A_{\text{amp}}/K)$ for different $\eta$ and $\zeta$.

The system amplifies the relative magnitude $A_{\text{amp}}/K$ as the driving frequency $\eta$ approaches the resonant state. In statistics, we rarely know ``where" to amplify. REAMP addresses this by designing a \textit{stochastic resonance system} at the expense of background noise. By introducing a randomized Beta density $f(x; \alpha, \beta)$---where $\alpha, \beta$ are independent Uniform variables from $U(a, b)$---we create a stochastic prior that allows every local structural peak in the data to resonate with a positive amplification probability.

This resonance logic shares a deep theoretical connection with \textit{Exponential Tilting Densities (ETD)}. As discussed in recent literature (e.g., \cite{SWWL24}), ETD minimizes the Kullback-Leibler (KL) divergence to approximate a baseline density under specific moment constraints. In our framework, the resonance step acts as a ``tilting"  mechanism that shifts the focus of the graph statistic toward latent structural breaks, providing a more robust normalization than the standard deterministic Ratio Cut $1/[i(n-i)]$. 

\subsection*{Graph Statistics and Double Sharpening}
The primary statistic used in REAMP is a graph-based count statistic derived from the \textit{Shortest Hamiltonian Path (SHP)}. Unlike the Minimum Spanning Tree (MST), the SHP possesses the properties of ``unboundedness" under the null hypothesis and ``boundedness" under the alternative \citep{Sh24}, ensuring high power in high dimensions. To maximize the geometric sensitivity of this graph, we utilize the \textit{Earth Mover’s Distance (EMD)} as our ground metric. EMD considers the minimum work required to transport feature mass between feature matrices $(X_i)$, allowing the model to detect changes in higher-order moments that Euclidean metrics would miss.
However, despite the success of REMAP built upon the SHP-based statistic, a direct extension of this statistic to the \textit{Sparsified Binary Segmentation (SBS)} approach \citep{CF15} fails in practice. This is because the REMAP framework generally returns only a single change point estimate after resonances, lacking the additive or separable structure required for the recursive localizations inherent in the SBS methodology.

Because the resonance system intentionally introduces stochasticity to ``vibrate"   the signal, the resulting distribution of estimates can be noisy. To extract the true change points, we implement a \textit{Double Sharpening} procedure. Based on the data sharpening iterations proposed by \cite{CBS22}, this iterative mean-shift process moves noisy data points toward the local modes of their own distribution. For $m = 0, 1,$ we set:
\begin{equation}\label{eqn:newiteration} 
y^{(m+1)}_i = y_i + y^{(m)}_i - \widehat{g}^{(m)}(x_i) 
\end{equation}
where $\widehat{g}^{(m)}(x)$ is the Nadaraya-Watson  estimator \citep{Na64, Wa64} applied to the sharpened data $y^{(m)}_i$. This sharpening acts as the functional opposite of REAMP; while resonance amplifies the signal across a broad search space, sharpening concentrates that amplified signal into precise, accurate modes representing the true change points. 

To finalize the estimation, we apply a localization step using two key \textit{R} functions. First, the \texttt{find.local.maxima} function is employed to identify the most significant resonance peak—filtering out background noise by applying a threshold of 0.05 to the sharpened density. This function isolates the continuous coordinate x where the statistical resonance is highest. Second, we apply the built-in \texttt{ceiling} function to this coordinate; this ensures the continuous peak location is mapped to the smallest following integer, providing a precise discrete time index for the estimated change point.

The remainder of this paper is organized as follows. In Section 2, we define the two-stage dimensionality reduction using EMD and the SHP count statistic, followed by the formal REAMP algorithm. Section 3 provides the asymptotic analysis of the resonance statistic. Section 4 presents numerical comparisons against state-of-the-art competitors. Section 5 demonstrates the potential of REAMP using a real-world cell image dataset, and we conclude with a discussion in Section 6.

\section{Methodology and Procedures}

\subsection{Two-Stage Dimension Reduction}

In high-dimensional image processing, the image histogram serves as a primary tool for dimensional reduction, offering both computational efficiency and invariance to object movement \citep{Bo09, PZM96}. For a sequence of high-dimensional data matrices $(Y_i)_{s \times d}$, we first represent each $Y_i$ by its relative occurrences across $p$ bins, denoted by the histogram vector $\bm{w}_i = (w_{i_1}, \ldots, w_{i_p})^\top$, where $\sum_{r=1}^p w_{i_r} = 1$. This transform reduces the $s \times d$ matrix to a $p$-dimensional vector.

However, histograms alone suffer from significant information loss, as vastly different images can share identical bin distributions \citep{Bo09}. To preserve the underlying structural signature, we propose a two-stage reduction. In the first stage, we extract a feature matrix $X_i \in \mathbb{R}^{p \times q}$, where the $(r,c)$-th element represents the $c$-th statistical feature (e.g., mean, variance, skewness) within the $r$-th bin. This ensures $q \ll p \ll \min\{s, d\}$.

In the second stage, we define the distance between these feature manifolds using the Earth Mover's Distance (EMD), which treats the problem as an optimal transport task \citep{Hi41}:
\begin{equation}\label{Earth}
c(i,j) = \min_{e(i_r, j_s) \geq 0} \sum_{r=1}^p \sum_{s=1}^p d(i_r, j_s) e(i_r, j_s),
\end{equation}
subject to the constraints:
\begin{equation*}
\sum_{s=1}^p e(i_r, j_s) = w_{i_r}, \quad \sum_{r=1}^p e(i_r, j_s) = w_{j_s}, \quad e(i_r, j_s) \geq 0.
\end{equation*}
Here, $d(i_r, j_s)$ is the ground distance (e.g., Manhattan or Euclidean) between the rows of $X_i$ and $X_j$. Unlike standard metrics, EMD incorporates the geometry of the bins, making it a superior measure for detecting shifts in variance and higher-order moments.

\textbf{Example 1 (Metric Sensitivity):} Let $p=2, q=1$ with equal weights $0.5$. If $X_i=(0.3, 0.7)^\top$ and $X_j=(0.4, 0.6)^\top$, the Manhattan distance is $0.2$, which is exactly double the EMD. More importantly, if $X_j$ is permuted to $(0.6, 0.4)^\top$, EMD remains invariant while the ground distance changes, demonstrating EMD's robustness to internal mass redistribution.

\textbf{Exampl/e 2 (Sensitivity to Variance Shifts).} To demonstrate the EMD's superior ability to capture distributional changes compared to standard Euclidean distance (ED), consider three normal random variables: $x \sim N(0, 0.5^2)$, $y \sim N(0, 0.2^2)$, and $z \sim N(0, 1.5^2)$. While all three share a common mean of zero, they are distinguished by their standard deviations: $0.5, 0.2,$ and $1.5$, respectively.

\begin{figure} [ht]
\centering
\makebox{\includegraphics[width=\textwidth]{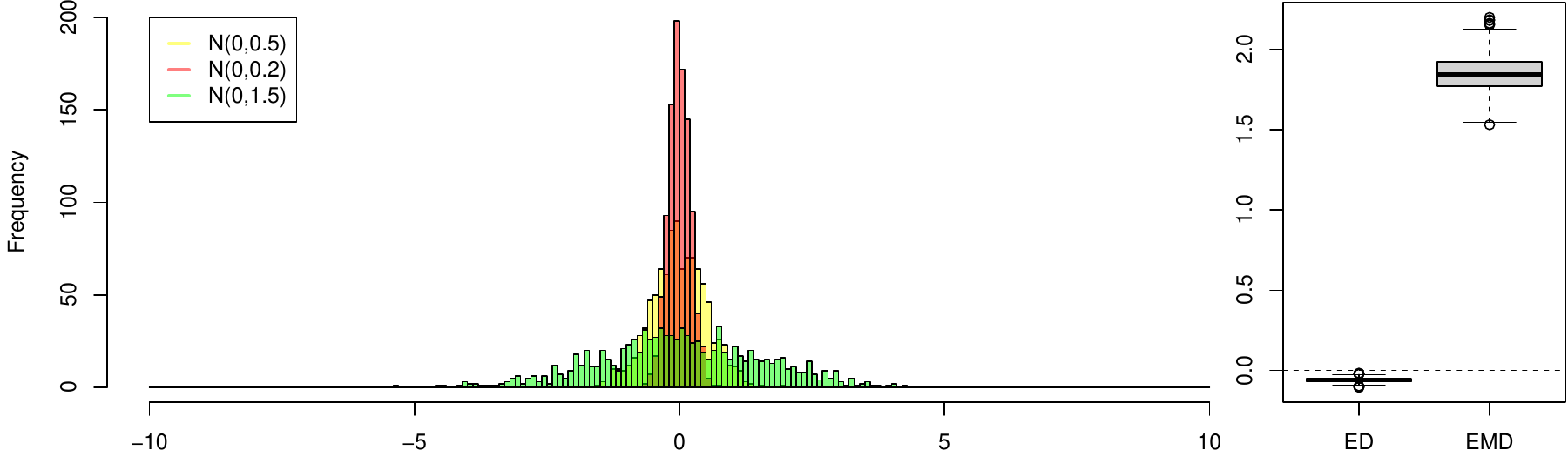}}
\caption{Comparisons between the ground Euclidean distance and the EMD with the ground Euclidean distance.}
\label{fig-hist}
\end{figure}

The left panel of Figure \ref{fig-hist} displays the histograms for 1,000 realizations of each variable within the range $[-10, 10]$. We evaluate the similarity between these distributions using both the standard ED and the EMD, utilizing the first two sample moments within each bin as features. The right panel of Figure \ref{fig-hist} illustrates the comparative results through the differences:
\begin{equation*}
\Delta_{ED} = d_{xz} - d_{xy} \quad \text{and} \quad \Delta_{EMD} = c_{xz} - c_{xy},
\end{equation*}
where $d$ and $c$ denote the derived ground Euclidean distance and EMD, respectively. 

Box plots generated from 1,000 independent repetitions reveal a stark contrast: $\Delta_{ED}$ is consistently negative, whereas $\Delta_{EMD}$ remains strictly positive. This indicates that while the standard Euclidean metric erroneously suggests $z$ is ``closer"   to $x$ than $y$, the EMD correctly identifies $y$ as the more similar distribution based on the standard deviation. 

These results confirm that the EMD is a significantly more robust similarity measure for detecting structural shifts in variance, which is a critical requirement for high-dimensional change point detection where mean-shifts may be absent but distributional shapes evolve.
 
Using EMD as the ground metric, we construct a global Similarity Graph via the Shortest Hamiltonian Path (SHP). The SHP $\hat{\P} = (\hat t_1, \ldots, \hat t_n)$ is found by solving the minimization:
\begin{equation}\label{Est1}
\hat{\P} = \arg\min_{\P \in \p_n} \sum_{i=1}^{n-1} c(t_i, t_{i+1}),
\end{equation}
where $\p_n$ is the set of all permutations. We utilize a heuristic Kruskal-based algorithm \citep{Sh23} to approximate $\hat{\P}$. The resulting count statistic $S_{\hat{\P}}(i)$ represents the number of edges in the SHP that cross the partition between the disjoint sets $\mathbb{N}_i = \{1, \ldots, i\}$ and $\bar{\mathbb{N}}_i$:
\begin{equation}\label{CS}
S_{\hat\P}(i) = \sum_{j=1}^{n-1} I\left[(\hat t_j \in \mathbb{N}_i \wedge \hat t_{j+1} \in \bar{\mathbb{N}}_i) \vee (\hat t_j \in \bar{\mathbb{N}}_i \wedge \hat t_{j+1} \in \mathbb{N}_i)\right].
\end{equation}

\subsection{Stochastic Resonance System}

While $S_{\hat{\P}}(i)$ is sensitive to structural breaks, traditional approaches often bias estimates toward the center of the sequence. To overcome this, we implement a stochastic resonance system. We define the resonance statistic $T_n(a, b)$ as:
\begin{equation}\label{cpe}
T_n(a, b) = \arg\min_{1 \leq i < n} \frac{S_{\hat\P}(i)}{f(i/n; \alpha, \beta)},
\end{equation}
where $\alpha, \beta \sim U(a, b)$ are independent random variables and $f(x; \alpha, \beta)$ is the Beta density function. In this framework, the Beta density acts as a ``stochastic prior" \citep{SWR22} that vibrates the system, allowing different change points to resonate as the global minimum across different iterations.

The count statistic $S_{\hat{\P}}(i)$ possesses two critical properties: it is \textit{unbounded} under a common distribution and \textit{bounded} under different distributions \citep{Sh24}. These properties, combined with the randomized denominator, mimic physical resonance, effectively amplifying subtle signals at any location in the series.

\begin{figure} [ht]
\centering
\makebox{\includegraphics[width=\textwidth]{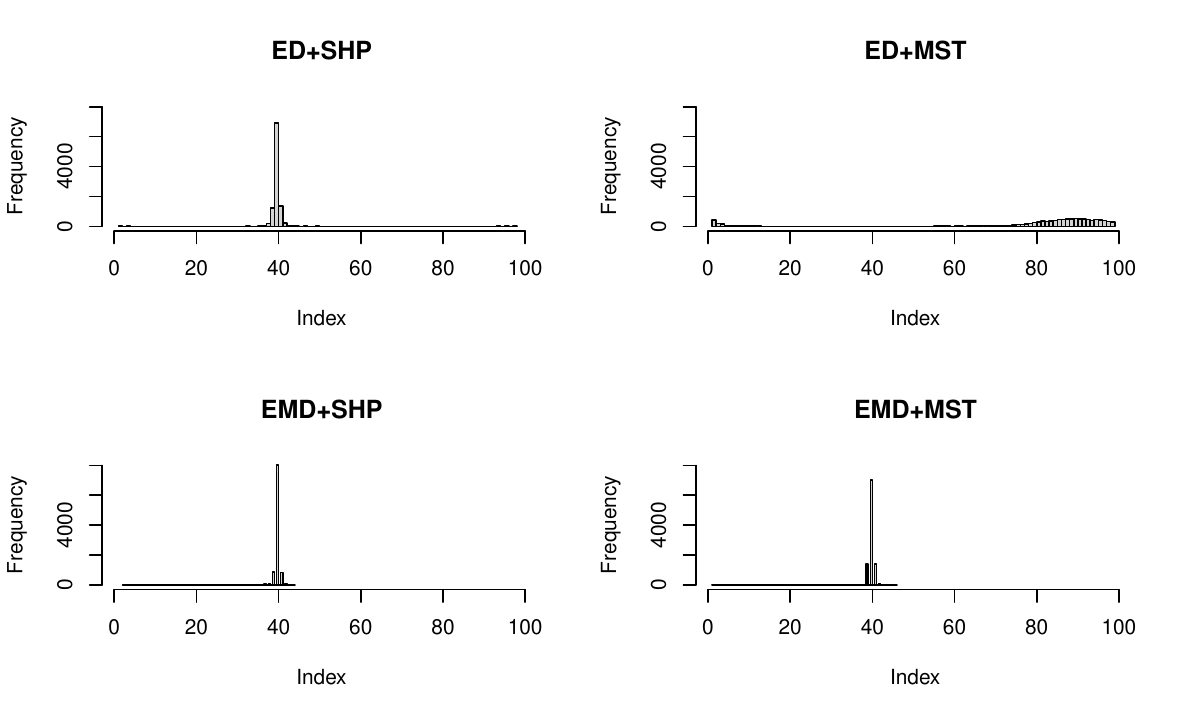}}
\caption{Comparison of REMAP statistics for detecting a variance shift. (Top Right) ED+MST fails, exhibiting spurious resonances. (Bottom Right) EMD improves MST performance, successfully resolving the peak. (Left Column) SHP-based statistics (ED and EMD) consistently isolate the change point with high precision, demonstrating superior power in high dimensions.}
\label{fig-mst-shp}
\end{figure}

We further examine the performance of REMAP in Example 2 by considering a change point at index 40 in a sequence of normal random vectors ($n=100$), where the variance shifts from $I_d$ to $5I_d$ with $d=1000$. We calculate   the resonated statistic $\frac{S_{\hat\P}(i)}{f(i/n; \alpha, \beta)}$ across 10,000 replications to generate the histograms displayed in Figure \ref{fig-mst-shp}. This scenario highlights the critical interplay between graph topology and the chosen ground metric in high-dimensional settings. While the ED+MST approach is insufficient for detecting the variance shift-yielding spurious peaks and failing to identify the true change point-the SHP-based statistics exhibit remarkable precision, producing a sharp spike at index 40 regardless of the metric. Furthermore, the transition to EMD as a ground metric serves as a stabilizing factor, enabling the MST-based statistic to overcome its inherent boundedness and correctly identify the structural break. Ultimately, the EMD+SHP configuration demonstrates the greatest geometric sensitivity, successfully isolating the change point with minimal statistical noise.

To resolve the noise introduced by the stochastic resonance, we apply a dual-stage sharpening procedure.  By iterating the resonance $L^+ = 10,000$ times, we generate a set of candidate points $\mathcal{T}$. The ``Double Sharpening" iteration is then applied to the estimated density  of $\mathcal{T}$. This process collapses the noisy resonance cloud into precise local modes, where we apply a 0.05 threshold to filter insignificant fluctuations, identify the peak coordinates via the \texttt{find.local.maxima} function, and utilize the built-in \texttt{ceiling} function to map these continuous values to the final discrete time index estimates. The complete process is detailed in Algorithm \ref{alg:REAMP}.

\begin{algorithm}[ht]
\caption{REAMP: Stochastic Resonance Sampling and Sharpening}
\label{alg:REAMP}
\begin{algorithmic}[1]
\State \textbf{Initialize:} $\mathcal{T} \gets \emptyset$, $L^+ \gets 10000$
\State Construct graph structure (SHP or MST) and compute count statistics $S_{\hat{P}}(i)$
\For{$L = 1$ \textbf{to} $L^+$}
    \State Sample resonance parameters $\alpha, \beta \sim U(a, b)$
    \State Identify candidate change point: $\tau^* \gets \arg\min_{1 \leq i < n} \left\{ \frac{S_{\hat{P}}(i)}{f(i/n; \alpha, \beta)} \right\}$
    \State Update resonance set: $\mathcal{T} \gets \mathcal{T} \cup \{\tau^*\}$
\EndFor

\State \textbf{Density Estimation:} Generate histogram from $\mathcal{T}$ with bin midpoints $X$ and densities $Y$
\State \textbf{Double Sharpening:} Apply iterative mean-shift via Eq. \eqref{eqn:newiteration} to obtain sharpened density $g^{(2)}$
\State \textbf{Local Maxima Extraction:}   Identify coordinates $x$ of local maxima in $g^{(2)}$ using \texttt{find.local.maxima} with threshold $0.05$
    \State \textbf{Localization:} Apply \texttt{ceiling} ($\cdot$) to map coordinates to discrete time indices
\State \Return $\widehat{\mathcal{T}}_{\text{CPS}}$ (Final localized change point estimates)
\end{algorithmic}
\end{algorithm} 

\section{Asymptotic Analysis}

     Let $g_{i}, g_{j}\in\{G_1,\ldots, G_{k^*+1}\} $ be the  samples containing $X_{i}$ and $X_{j}$, respectively.   Let the leading term  of a Taylor series of the cost $c(i, j)$ about some non-random point  of interest be $c^*(g_{i}, g_{j})$.  Assume that 
$c(i, j)-c^*(g_{i}, g_{j})$ has some convergence rate.  To establish the theoretical framework, we make the following assumptions.

{\bf A1.} $c(i, j)>0, i\neq j,~\text{almost~ surely.}$

{\bf A2.}
$c(i, j)=c(j, i) ~(\text{symmetry}).$

{\bf A3.}
$c(i, j)\leq c(i, \bar{i})+c(\bar{i}, j) ~(\text{triangular~ inequality}).$

{\bf A4.}
 For any small $\varepsilon>0$, $P(|c(i, j)-c^*(g_{i}, g_{j})|>\varepsilon)=O(\eta_{p,q})$ and  $n^2\eta_{p,q}\rightarrow0$ (rates) as $p\rightarrow\infty$, where $p$ is the number of bins.

{\bf A5.}
If $F_{g_i}\neq  F_{g_j}, ~\text{then}~c^*(g_i, g_j)>\min\{c^*(g_i, g_i),c^*(g_j, g_j)\}$ (increased cost).

{\bf A6.}  $\min_{\ell=1,\ldots,k^*}\tau_\ell/n>0$.

{\bf A7.} $\min_{\ell=1,\ldots,k^*}f(\tau_\ell/n;\alpha,\beta)>0$.

{\bf A8.}  $P\left\{f(\tau_{\ell_1}/n; \alpha_1, \beta_1)> 2f(\tau_{\ell_2}/n; \alpha_2, \beta_2)\right\}>0$ for all $\ell_1\neq \ell_2\in \{1,\ldots,k^*\}$.

A1-A5 are adjusted  from \cite{Sh24} in terms of EMD. As shown in \cite{RTG00}, the EMD is a metric.  Thus, if we require the data distribution to be continuous, then A1-A3 hold. A4 requires that the dimension of bins $p$ should be higher than $n$ or less than $n$ if the data are light-tailed. In A5, We require  that the between-sample EMD must be greater than one of within-sample EMD.  See \cite{Sh24} for more details.  A6 assumes that the distance between any two neighboring change points should not be too close.
A7 requires that the Beta density $f(x; a,b)$ at any change point should not be close to zero, which is weaker than its boundedness from below. A8 requires that the resonance system be properly designed, where the number 2 is derived from the boundedness property of the count statistic in Lemma 1.

{\bf Lemma 1 (Two properties).}
Assume that  A1-A5 hold. The following two properties hold:
\begin{itemize}
 
\item[1.] (Boundedness) $S_{\hat\P}(\tau_\ell)\leq 2$ in probability for $\ell=1,\ldots,k^*$.
\item[2.] (Unboundedness) $\frac{1}{n}S_{\hat\P}(i)\geq\gamma_1$ if $|i-\tau_{\ell}|/n>\gamma_2$ for some positive constants $\gamma_1$ and $\gamma_2$, and   $\ell=1,\ldots,k^*$.
\end{itemize}

The proof of Lemma 1 is a direct result of \cite{Sh24}. Next, we provide our theoretical result in Theorem 1.
 
{\bf Theorem 1.}  Assume that A1-A8 hold. For any small $\epsilon>0$,
$$P(\min_{\ell\in\{1,\ldots,k^*\}}|T_n(a, b)/n-\tau_\ell/n|<\epsilon)\rightarrow1$$ $$\text{and}~ P(|T_n(a, b)/n-\tau_\ell/n|<\epsilon)>0, ~\text{for any}~\ell\in\{1,\ldots,k^*\}.$$

According to Theorem 1, the estimated change point $T_n(a, b)$ can be a neighborhood of any  change point.


\section{Numerical Simulations}

To evaluate the performance of the proposed REAMP method, we conduct a series of numerical simulations. While REAMP is designed for general matrix-valued data through histogram transformation, most competing methods are optimized for vector-valued inputs. Consequently, we consider a sequence of vectors $Y_i \in \mathbb{R}^d$ ($s=1$). The components of $Y_i$ are partitioned into two halves: the first $d/2$ components are drawn from a normal distribution $N(\mu_{i,1}, \sigma_{i,1}^2)$, and the remaining $d/2$ components are drawn from $N(\mu_{i,2}, \sigma_{i,2}^2)$. The parameter vectors $(\mu_{i,1}, \sigma_{i,1}, \mu_{i,2}, \sigma_{i,2})$ are defined piecewise as follows:
\begin{align}\label{pars}
(\mu_{i,1}, \sigma_{i,1}, \mu_{i,2}, \sigma_{i,2}) = \begin{cases} 
(1, 1, 1, 1) & \text{for } 1 \leq i \leq \tau_1 \\
(1, 1, -1, 1) & \text{for } \tau_1 < i \leq \tau_2 \\
(-1, 2, -1, 2) & \text{for } \tau_2 < i \leq \tau_3 \\
(0, 1, 0, 1) & \text{for } \tau_3 < i \leq n.
\end{cases}
\end{align}
This setup incorporates $k^*=3$ structural change points involving shifts in both mean and variance. We investigate two distinct scenarios to test the scalability of the method:
\begin{itemize}
    \item \textbf{Scenario 1:} $n=100$, $d=200$ with true change points at $(\tau_1, \tau_2, \tau_3) = (20, 40, 75)$.
    \item \textbf{Scenario 2:} $n=200$, $d=400$ with true change points at $(\tau_1, \tau_2, \tau_3) = (40, 80, 150)$.
\end{itemize}

\begin{figure} [ht]
\centering
\makebox{\includegraphics[width=\textwidth]{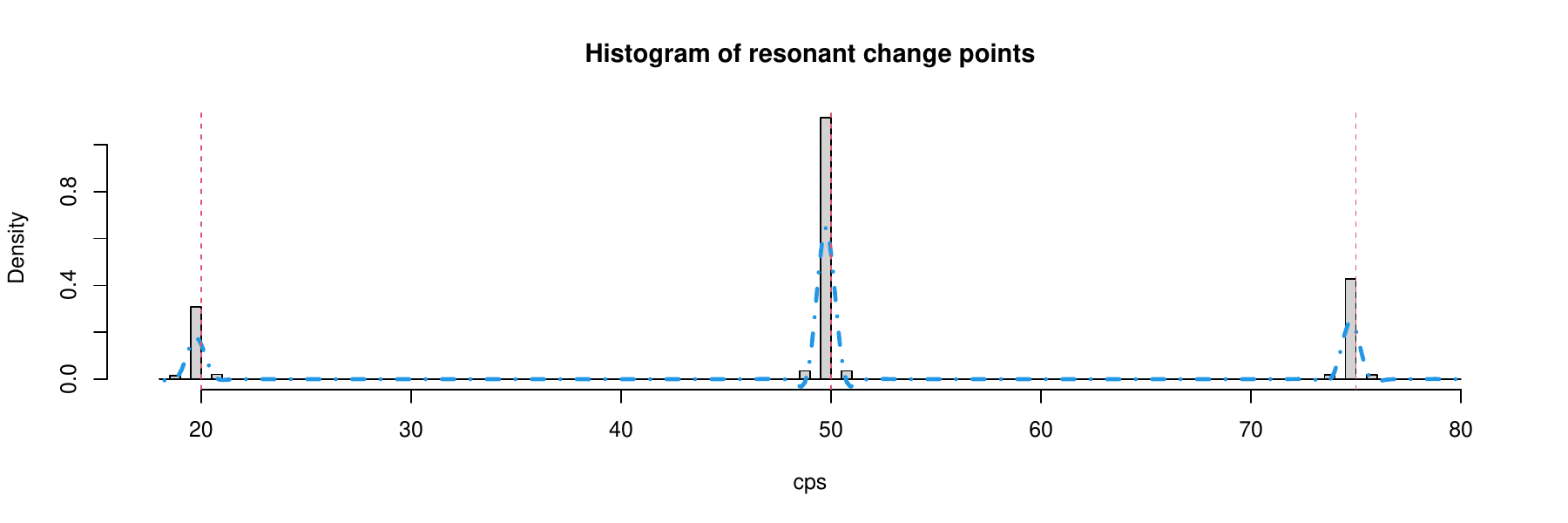}}
\caption{ 
Histogram and sharpening curve of $T_n(a, b)$ and estimated change points at 20, 50, and 75.}
\label{fig-simu}
\end{figure}

For Scenario 1, we set the number of initial bins $nb_1 = 100$ and the secondary bins for resonance $nb_2 = 90$. For Scenario 2, $nb_1 = 100$ and $nb_2 = 180$ were selected to enhance the filtering effect for larger sample sizes. Figure \ref{fig-simu} illustrates the raw resonance distribution and the resulting sharpened density curve for a single realization of Scenario 1. In this specific instance, the REAMP method successfully identifies the change points at $20, 50,$ and $75$.

We repeat each simulation scenario $200$ times. The accuracy of the estimated change point set $\mathcal{A}$ relative to the true set $\mathcal{B} = \{\tau_1, \tau_2, \tau_3\}$ is quantified using the Hausdorff distance:
\begin{equation}
H(\mathcal{A}, \mathcal{B}) = \max \{ h(\mathcal{A}, \mathcal{B}), h(\mathcal{B}, \mathcal{A}) \},
\end{equation}
where $h(\mathcal{A}, \mathcal{B}) = \max_{a \in \mathcal{A}} \min_{b \in \mathcal{B}} |a - b|$. This metric penalizes both the failure to detect true change points and the inclusion of false positives.

We compare REAMP against two state-of-the-art high-dimensional detection methods: the E-statistic method by \cite{MJ14} (MJ) and the \textit{Inspect} method by \cite{WS18} (WS). Table \ref{table1} reports the average Hausdorff distance (AH) and the average estimated number of change points (AN).

\begin{table}[!h]
\centering
\caption{Comparative performance of REAMP, MJ, and WS over 200 repetitions. AH denotes average Hausdorff distance; AN denotes average number of detected change points (True $k^*=3$).}
\label{table1}
\begin{tabular}{lcccc} 
\hline
& \multicolumn{2}{c}{\textbf{Scenario 1} ($n=100$)} & \multicolumn{2}{c}{\textbf{Scenario 2} ($n=200$)} \\ 
\cline{2-5}
\textbf{Method} & AH & AN & AH & AN \\ \hline
WS \citep{WS18} & 17.00 & 23.69 & 34.65 & 44.32 \\
MJ \citep{MJ14} & 20.00 & 2.00 & 1.16 & 3.04 \\
REAMP           & \textbf{1.55} & \textbf{3.10} & \textbf{1.46} & \textbf{3.09} \\ 
\hline
\end{tabular}
\end{table}

The results in Table \ref{table1} demonstrate that REAMP provides consistently accurate estimations across both scenarios, maintaining an estimated number of change points (AN) close to the ground truth. In contrast, WS significantly overestimates the number of change points; this is likely because WS is optimized for sparse mean-shifts and struggles when faced with simultaneous shifts in mean and variance. The MJ method performs well in the larger sample size of Scenario 2 but fails in Scenario 1, suggesting that MJ requires a larger ratio of $n$ to $d$ to achieve stability, whereas REAMP's graph-based resonance remains robust in higher-dimensional, small-sample settings.

  \section{Real Data Application: Cell Division Analysis}

Analyzing the temporal distribution of cell divisions is critical for assessing the health and viability of biological organisms. Deviations in division rates—either accelerated or suppressed—can serve as early indicators of pathological conditions, such as oncogenesis or developmental abnormalities, where intervention or process termination may be required. 

In this application, we employ the REAMP framework to identify structural breaks in a time-lapse image sequence obtained from an automated embryo monitoring system \citep{CGGG14}. The dataset consists of $n = 285$ frames, each with a spatial resolution of $321 \times 321$ pixels.

Following the parameter selection guidelines established by \cite{SWR17}, we set the initial number of bins to $nb_1 = 50$. To test the sensitivity and robustness of the stochastic resonance system, we varied the secondary bin parameter $nb_2$ across the range $\{50, 60, 70, 80, 90\}$. Regardless of the $nb_2$ configuration, the REAMP method consistently identified three distinct change points at frames 23, 192, and 268. 

\begin{figure} [ht]
\centering
\includegraphics[width=\textwidth]{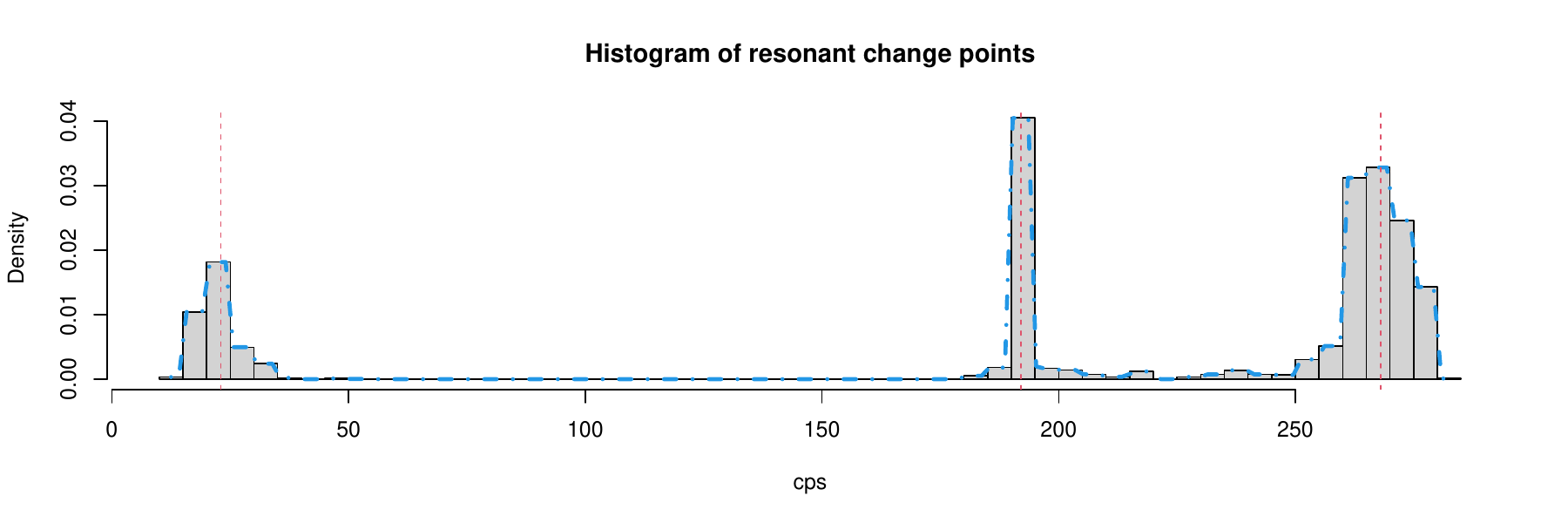}
\caption{Histogram and sharpening curve for $T_n(a, b)$ with estimated change points identified at frames 23, 192, and 268.}
\label{fig-cell}
\end{figure}

Figure \ref{fig-cell} illustrates the histogram of resonant candidate points for the $nb_1 = nb_2 = 50$ configuration, alongside the corresponding sharpening curves. These curves clearly delineate the three estimated change points, highlighting the precision of the resonance-sharpening procedure.

While our results are numerically consistent with the findings of \cite{SWR17}, the REAMP approach offers a significant advantage: it provides a continuous, density-based visualization of change-point locations. Unlike the binary segmentation method utilized in \cite{SWR17}, which provides point estimates through discrete recursive partitions, REAMP’s resonance cloud and sharpened density offer an intuitive mapping of the statistical confidence and structural signal strength throughout the entire division process.

\section{Conclusion and Discussion}

In this paper, we introduced the \textbf{REAMP} framework, a novel methodology for the detection and visualization of multiple change points in high-dimensional data. By integrating the geometric sensitivity of the Earth Mover's Distance (EMD) with a stochastic resonance system, REAMP transforms the discrete problem of change point estimation into a continuous density-estimation task. 

The conceptual foundation of our approach—the resonance system—represents a significant interdisciplinary bridge between statistical signal processing and classical physics. Such cross-pollination of knowledge offers unique perspectives on data volatility and structural stability that are often overlooked in traditional statistical frameworks. Our simulation studies demonstrate that REAMP remains robust under high-dimensional settings where shifts in both mean and variance occur simultaneously, consistently outperforming established methods such as \textit{Inspect} and the E-statistic-based \textit{ecp}.

The application to real-world time-lapse image data in cell division analysis further underscores the practical utility of our method. Beyond providing accurate numerical estimates, REAMP offers a visual representation of the ``resonance'' at each candidate point, allowing researchers to qualitatively assess the strength and clarity of structural breaks across the temporal domain. 

While our current theoretical framework is built upon graph-based count statistics, the principles of stochastic resonance are generalizable. Future research will explore the extension of these resonance-based priors to traditional CUSUM-style statistics and likelihood-ratio frameworks. Furthermore, investigating the optimal selection of the Beta-density hyperparameters ($a, b$) through adaptive learning remains a promising avenue for improving detection sensitivity in extremely noisy environments.
  
\section{Acknowledgments}

We thank Dr. M. Cicconet for sharing the cell image data. We thank Dr. Jiahua Chen for the helpful discussion on the Earth Mover's Distance. Li's Research was partially supported by NSFC (Grant
No. 12271047) and Guangdong and Hong Kong Universities “1+1+1” Joint Research Collaboration Scheme (2025A0505000010).  Shi was supported by the Natural Sciences and Engineering Research
Council of Canada under Grant RGPIN-2022-03264, the Interior Universities Research Coalition
and the BC Ministry of Health, the NSERC Alliance International Catalyst Grant ALLRP
590341-23, and the University of British Columbia Okanagan (UBC-O) Vice Principal Research in collaboration with UBC-O Irving K. Barber Faculty of Science.

 \end{document}